\documentclass[10pt,conference]{IEEEtran}
\IEEEoverridecommandlockouts

\usepackage{graphics} 
\usepackage{epsfig}
\usepackage{mathptmx} 
\usepackage{times} 
\usepackage{amsmath} 
\usepackage{amssymb}

\usepackage{cite}
\usepackage{balance}
\usepackage[amsmath,amsthm,thmmarks]{ntheorem}
\usepackage{amsfonts}
\usepackage{mathrsfs}
\usepackage{mathtools}
\usepackage[dvipsnames]{xcolor}
\usepackage{enumerate}
\usepackage{enumitem}
\usepackage[bbgreekl]{mathbbol}
\usepackage{colortbl}
\usepackage{relsize}
\usepackage{etoolbox}
\usepackage{multirow}
\usepackage{adjustbox}
\usepackage{makecell}
\usepackage{tabularx,ragged2e,booktabs}
\newcolumntype{R}[1]{>{\RaggedLeft\arraybackslash}p{#1}}
\usepackage{tabularx, booktabs, makecell}
\usepackage{siunitx}
\usepackage{standalone}
\usepackage{tikz}
\usepackage{pgfplots}
\pgfplotsset{compat=1.6, scaled x ticks = false, 
xticklabel style={/pgf/number format/fixed,/pgf/number format/precision=3},
legend image post style={xscale=0.5},
legend image post style={yscale=0.6}} 
\usepackage{bbold}
\usepackage{float}
\usepackage{hyperref}
\usepackage{blkarray}
\usepackage{graphicx}
\usepackage{tikz}
\usepackage{pgfplots} 
\usepackage{pgfgantt}
\usepackage{pdflscape}
\usepgfplotslibrary{units}
\newlength\fheight 
\newlength\fwidth 
\usetikzlibrary{decorations.pathreplacing,calligraphy}
\usetikzlibrary{pgfplots.groupplots}
\usepackage{framed}
\usetikzlibrary{matrix}

\newcolumntype{P}[1]{>{\centering\arraybackslash}p{#1}}
\newcolumntype{C}[1]{>{\centering\arraybackslash}m{#1}}
\newcolumntype{L}[1]{>{\raggedleft\arraybackslash}m{#1}}
\newcolumntype{s}{>{\hsize=.8\hsize}X}

\newcommand{\ols}[1]{\mskip.5\thinmuskip\overline{\mskip-.5\thinmuskip {#1} \mskip-.5\thinmuskip}\mskip.5\thinmuskip} 
\usetikzlibrary{decorations.pathreplacing,calligraphy}

\makeatletter
\newcommand{\linebreakand}{%
  \end{@IEEEauthorhalign}
  \hfill\mbox{}\par
  \mbox{}\hfill\begin{@IEEEauthorhalign}
}
\makeatother

\begin{document}

\title{A Fast Dynamic Internal Predictive Power Scheduling Approach for Power Management in Microgrids\\

\thanks{This work was authored by the National Renewable Energy Laboratory, operated by Alliance for Sustainable Energy, LLC, for the U.S. Department of Energy (DOE) under Contract No. DE-AC36-08GO28308. Funding is provided by NREL Laboratory Directed Research and Development Program. The views expressed in the article do not necessarily represent the views of the DOE or the U.S. Government. The U.S. Government retains and the publisher, by accepting the article for publication, acknowledges that the U.S. Government retains a nonexclusive, paid-up, irrevocable, worldwide license to publish or reproduce the published form of this work or allow others to do so, for the U.S. Government purposes.}
}

\author{
\IEEEauthorblockN{Neethu Maya}
\IEEEauthorblockA{Department of Aerospace Engineering\\
\emph{Indian Institute of Science}\\
Bangalore, India\\
neethum@iisc.ac.in}
\and
\IEEEauthorblockN{Bala Kameshwar Poolla}
\IEEEauthorblockA{\emph{National Renewable Energy Laboratory}\\
Golden CO, USA\\
bpoolla@nrel.gov}
\and
\IEEEauthorblockN{Seshadhri Srinivasan}
\IEEEauthorblockA{\emph{TVS Sensing Solutions}\\
Madurai, India\\
seshucontrol@gmail.com}
\linebreakand
\IEEEauthorblockN{Narasimhan Sundararajan}
\IEEEauthorblockA{School of Electrical and Electronics Engineering\\
\emph{Nanyang Technological University}\\ Singapore\\
ensundara@gmail.com}
\and
\IEEEauthorblockN{Suresh Sundaram}
\IEEEauthorblockA{Department of Aerospace Engineering\\
\emph{Indian Institute of Science}\\
Bangalore, India\\
vssuresh@iisc.ac.in}
}

\maketitle
\vspace{-2cm}
\begin{abstract}
This paper presents a Dynamic Internal Predictive Power Scheduling (DIPPS) approach for optimizing power management in microgrids, particularly focusing on external power exchanges among diverse prosumers. DIPPS utilizes a dynamic objective function with a time-varying binary parameter to control the timing of power transfers to the external grid, facilitated by efficient usage of energy storage for surplus renewable power. The microgrid power scheduling problem is modeled as a mixed-integer nonlinear programming (MINLP-PS) and subsequently transformed into a mixed-integer linear programming (MILP-PS) optimization through McCormick’s relaxation to reduce computational complexity. A predictive window with 6 data points is solved at an average of 0.92s, a 97.6\% improvement over the 38.27s required for the MINLP-PS formulation, implying the numerical feasibility of the DIPPS approach for real-time implementation. Finally, the approach is validated against a static objective using real-world load data across three case studies with different time-varying parameters, demonstrating the ability of DIPPS to optimize power exchanges and efficiently utilize distributed resources while shifting the external power transfers to a specified time duration.
\end{abstract}


\section{Introduction}
Microgrids are electrical systems with local distributed power generation e.g., photovoltaic (PV) installations integrated with electrical storage systems (ESS) \cite{denholm_evaluating_2007}, that can operate independently (stand-alone) or in conjunction with the external utility grid. Internal power scheduling of microgrids involves the optimal scheduling of distributed power generation systems and electrical storage. It also requires external power exchanges with utility grids or peers to minimize power curtailment while maximizing sustainable resource utilization and revenues. 

The microgrid ESS integration approaches depend on numerous factors such as total load, generation, cost of procurement, efficiency, and scalability \cite{emrani_comprehensive_2024}. The sustainability of energy consumption and accrued cost savings depend on the system objectives of the microgrid power scheduling schemes. A cost-benefit analysis for multi-energy microgrids using mixed-integer linear programming (MILP) optimization is presented in \cite{shahbazbegian_techno_economic_2023}. Further, the advantages of ESS in microgrids, such as reliability in power availability, energy time-shifting, and arbitrage through time-of-use (ToU) tariff optimization, are discussed in \cite{wicke_hierarchical_2024}. Additionally, a thorough comparison of existing power scheduling methods is available in \cite{azuatalam_energy_2019}, while \cite{akter_review_2024} reviews previous studies using meta-heuristic approaches for power scheduling.

The ESS charging/discharging patterns, rates, and states of charge (SoC) are heavily influenced by the power scheduling in microgrids \cite{xue_continuous_2024}. While many existing methods focus on short-term revenues \cite{azuatalam_energy_2019}, they often overlook the long-term impacts on the ESS lifespans. The economic viability and profitability of microgrids \cite{shabani_smart_2024} are inherently tied to the storage lifespans (typically $10$ years for lead-acid batteries \cite{akter_optimal_2024}). It is, therefore, crucial to accurately model storage systems. Factors such as charging/discharging durations, peak currents, and cycle counts critically affect the ESS lifespan. Limiting the SoC to lower levels too can extend the storage life and increase overall revenue \cite{perez_effect_2016}, especially for batteries with a shorter cycle life (e.g., lead-acid batteries)\cite{fares_what_2018}.

Government policies such as time-of-use and feed-in tariffs encourage reduced grid dependence and increase profitability through peer-to-peer (P2P) power trading schemes \cite{dusonchet_comparative_2015}. Recent studies on revenue generation from external power exchanges have highlighted the potential of P2P trading between interconnected microgrids. Developing power scheduling models that consider ESS lifespan and can simulate large-scale microgrid communities is essential for exploring P2P power exchange dynamics. The heterogeneity in distributed power generation, storage, and load scheduling among peers is crucial for efficient and cost-effective P2P exchanges \cite{chen_holistic_2024}. Thus, a comprehensive power scheduling model incorporating ESS modeling complexities and reflecting the diverse decision-making of individual microgrid prosumers is necessary.

To address these challenges, a Dynamic Internal Predictive Power Scheduling (DIPPS) approach is presented in this paper. DIPPS enables the simulation of scalable, heterogeneous microgrid communities with distributed power generation, providing insights into optimizing P2P power exchanges. At the outset, the internal power scheduling is modeled as a Mixed Integer Non-Linear Programming (MINLP) optimization with Predictive Scheduling (PS). To enable real-time implementation, the MINLP problem is reduced and reformulated as a Mixed Integer Linear Programming (MILP)-PS problem. Further, in this MILP-PS problem, the objective is modified dynamically by including a time-varying parameter that can be adjusted to optimize internal power scheduling and control heterogeneous external power exchanges. Detailed performance evaluations show that the DIPPS approach is also capable of shifting the timing of power exchanges. For the ESS systems, a simplified linear formulation is used where the charge-discharge cycles are limited to one per day by adjusting the predictive scheduling window and power capacities. Three distinct scenarios with varying PV, ESS, and load capacities are considered for simulations to evaluate the scope of the approach. 

The remainder of this paper is organized as follows: Section \ref{Sec_Formulation} introduces the microgrid internal power scheduling framework followed by the proposed DIPPS approach. Next, Section \ref{Sec_Results} details the performance evaluation and presents simulation results for the DIPPS approach. Finally, Section \ref{Sec_Conclusion} summarizes the key findings of this study and outlines potential avenues for future research.

\section{Dynamic Internal Predictive Power Scheduling Problem Formulation}
\label{Sec_Formulation}
\subsection{Modeling preliminaries}

\begin{figure}[!htbp]	
    \centering
    \usetikzlibrary{calc,arrows}
\usetikzlibrary{positioning}
\usetikzlibrary{scopes,patterns,intersections,calc}
\begin{tikzpicture}[x=2.9in,y=3.5in]
   
\draw[line width=0.25mm] (0.14,0.25) rectangle (0.42,0.35);
\node[fill=white, inner sep=0.3ex] at (0.275,0.30) {\small \shortstack[c]{PV\\$P^{PV}(t)$}};
			
\node[trapezium,draw,trapezium stretches = true,line width=0.25mm,shape border rotate=180,
minimum width = 1.9, minimum height = 0.13,trapezium angle = 80] (t) at (0.70,0.40) {\small \shortstack[c]{EL\\$P_{L}(t)$}};
			
\draw[line width=0.25mm] (0.15,0.45) rectangle (0.39,0.58);
\node[fill=white, inner sep=0.3ex] at (0.27,0.515) {\small \shortstack[c]{ESS\\$SoC(t)$}};
			
\draw[line width=0.25mm] (0.18,0.70) circle (6pt);
\draw[line width=0.25mm] (0.36,0.70) circle (6pt);
\draw[line width=0.25mm] (-0.01,0.75) rectangle (0.86,0.21);
\draw[line width=0.25mm] (-0.01,0.85) rectangle (0.86,0.92);

\node[fill=white, inner sep=0.3ex] at (0.435,0.885) {\small \shortstack[c]{Utility Grid and P2P exchange}};
\draw[line width=0.25mm,-to] (0.18,0.725) -- (0.18,0.85);
\node[fill=white, inner sep=0.3ex] at (0.16,0.80) {\small \shortstack[c]{$P^{S}_{G}(t)$}};			
\draw[line width=0.25mm,-to] (0.36,0.85) -- (0.36,0.725);

\node[fill=white, inner sep=0.1ex] at (0.35,0.80) {\small \shortstack[c]{$P^{G}_{B}(t)$}};

\draw[line width=0.25mm,-to] (0.14,0.3) -- (0.07,0.3) |- (0.15,0.7);
\node[fill=white, inner sep=0.3ex] at (0.1,0.4) {\small \shortstack[c]{$P^{PV}_{G}(t)$}};

\draw[line width=0.25mm,-to] (0.27,0.35) -- (0.27,0.45);
\node[fill=white, inner sep=0.3ex] at (0.29,0.39) {\small \shortstack[c]{$P^{PV}_{ES}(t)$}};

\draw[line width=0.25mm,-to] (0.18,0.58) -- (0.18,0.675);
\node[fill=white, inner sep=0.3ex] at (0.19,0.62) {\small \shortstack[c]{$P^{ES}_{G}(t)$}};		

\draw[line width=0.25mm,-to] (0.36,0.675) -- (0.36,0.58);
\node[fill=white, inner sep=0.1ex] at (0.37,0.63) {\small \shortstack[c]{$P^{G}_{ES}(t)$}};		

\draw[line width=0.25mm,-to] (0.39,0.515) -| (0.65,0.449);
\node[fill=white, inner sep=0.3ex] at (0.5,0.515) {\small \shortstack[c]{$P^{ES}_{L}(t)$}};			
\draw[line width=0.25mm,-to] (0.39,0.7) -| (0.74,0.449);
\node[fill=white, inner sep=0.3ex] at (0.74,0.6) {\small \shortstack[c]{$P^{G}_{L}(t)$}};			
\draw[line width=0.25mm,-to] (0.42,0.3) -| (0.70,0.35);
\node[fill=white, inner sep=0.3ex] at (0.54,0.3) {\small \shortstack[c]{$P^{PV}_{L}(t)$}};

\end{tikzpicture}
    \caption{Microgrid power flow schematic}
    \label{Fig_ELModel}
\end{figure}
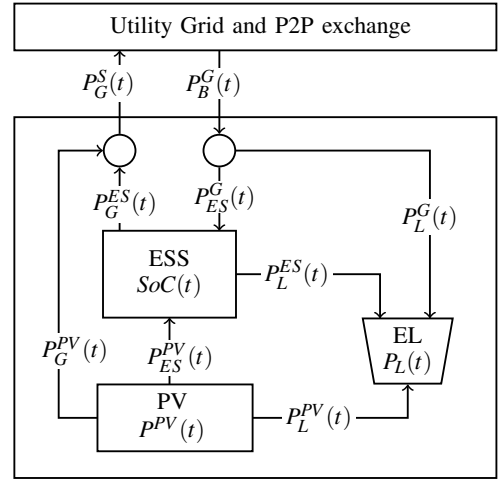

A microgrid with PV and ESS as shown in Fig.~\ref{Fig_ELModel} is considered in this paper. The power flow variables used in this paper are denoted as $P^{A}_{B}$ to represent the power transferred from source $A$ to destination $B$. $P_{L}(t)$ denotes the total electrical load, $P^{PV}(t)$ the power produced by PV, and $SoC(t)$ the state of charge of the ESS. $P^{G}_{S}(t)$ and $P^{B}_{G}(t)$ are used to denote the total external power exchange, which includes both power exchange with the utility grid and the P2P exchange within the community. 

To ensure mutual exclusivity of power flows, three binary variables are introduced. The variable $b_{V}(t)$ ensures that the charging of the ESS from the external grid $P^{G}_{ES}(t)$ and charging from PV, $P^{PV}_{ES}(t)$ are mutually exclusive. This helps in minimizing any losses incurred in DC-AC conversions. The purchasing and selling to/from the external grids ($P_{B}^{G}(t), P_{G}^{S}(t) ~\text{resp.}$) are mutually exclusive through the binary variable $b_{G}(t)$. Finally, the variable $b_{C}(t)$ provides for mutual exclusivity of the charging and discharging ($P_{ES}^{C}(t), P^{ES}_{D}(t) ~\text{resp.}$) of the ESS. These three binary variables are defined as
	\begin{align}
		\label{binVars_01}
		&	b_{V}(t)\coloneqq 
		\begin{cases}
			0 & P^{G}_{ES}(t) \geq 0\\
			1 & P^{PV}_{ES}(t) \geq 0,
		\end{cases}\notag
		,\,	b_{G}(t) \coloneqq 
		\begin{cases}
			0 & P_{G}^{S}(t) \geq 0 \\
			1 & P_{B}^{G}(t) \geq 0,
		\end{cases}\notag\\
		&  b_{C}(t) \coloneqq 
		\begin{cases}
			0 & P^{ES}_{D}(t) \geq 0\\
			1 & P_{ES}^{C}(t) \geq 0.			
		\end{cases}
	\end{align}

 The electrical load at time instant $t$, $P_{L}(t)$ is met from the power purchased from grid $P^{G}_{L}(t)$, the discharge from the ESS $P^{ES}_{L}(t)$, and the PV production $P^{PV}_{L}(t)$ i.e.,
	\begin{equation}
		\label{EQ_ELCons}
		P_{L}(t) = P^{G}_{L}(t)b_{G}(t) + P^{ES}_{L}(t)(1-b_{C}(t)) + P^{PV}_{L}(t).
	\end{equation}
 
Similarly, the PV generated power $P^{PV}(t)$ is used to either meet the load through $P^{PV}_{L}(t)$, charge the ESS as $P^{PV}_{ES}(t)$, or is sold to the external grid as $P^{PV}_{G}(t)$, i.e.,
	\begin{align}
		\label{EQ_SolarCons}
		P^{PV}(t) = &~P^{PV}_{G}(t)(1-b_{G}(t)) + P^{PV}_{L}(t) \\\notag+ & P^{PV}_{ES}(t)b_{V}(t)b_{C}(t).
	\end{align}
 
  Likewise, the ESS is charged directly from the external grid as $P^{G}_{ES}(t)$ or from the PV as $P^{PV}_{ES}(t)$. The ESS discharges to either meet the load demand as $P^{ES}_{L}(t)$ or sell back to the external grid as $P^{ES}_{G}(t)$. These are represented by
	\begin{align}
		\label{EQ_Power_CH_DIS}
		P_{ES}^{C}(t) = &~P^{G}_{ES}(t)b_{G}(t)(1-b_{V}(t))+P^{PV}_{ES}(t)b_{V}(t),\notag \\
		P^{ES}_{D}(t) = &~P^{ES}_{G}(t)(1-b_{G}(t))+P^{ES}_{L}(t).
	\end{align}
 
The change in the state of charge $SoC(t)$ of the ESS depends on the charging power $P_{ES}^{C}(t)$ and the discharging power $P^{ES}_{D}(t)$. Throughout this paper, the state of charge is constrained as $0.2 \leq SoC(t) \leq 0.8$ to avoid overcharging and complete draining of the ESS. Furthermore, let $\eta_C$ and $\eta_D$ be the charging and discharging efficiencies, then the variation of the SoC of the ESS is given by
	\begin{align}
		\label{EQ_DES_SoC}
		SoC(t+1)=&~SoC(t) - (\eta_D\,P^{ES}_{D}(t)\,(1-b_{C}(t))\notag \\+ &\eta_C\,P_{ES}^{C}(t)\,b_{C}(t))/ES_{cap},
	\end{align}
where $ES_{cap}$ is the capacity of the ESS in each microgrid. The initial SoC of the ESS is fixed at $0.5$ and the final SoC is also constrained to be $0.5$ such that the ESS can provide power in the absence of PV. The charging and discharging power of the ESS are also constrained in terms of $ES_{cap}$ (the maximum charging power is $40\%$ of $ES_{cap}$ and discharging power is set at $20\%$ of $ES_{cap}$) i.e.,
	\begin{align}
		\label{EQ_DES_P_chmax}
		&	0 \leq P_{ES}^{C}(t) \leq \ols{P^{ES}_{C}},\,~
		0 \leq P^{ES}_{D}(t) \leq \ols{P^{ES}_{D}},\\\notag
		&\ols{P^{ES}_{C}} = 0.4\times ES_{cap},\,~
		\ols{P^{ES}_{D}} = 0.2\times ES_{cap}.
	\end{align}

Finally, the total external power exchanges from the microgrid at time $t$ is given by,
	\begin{align}
		\label{EQ_UG_ex}
		P^{G}_{B}(t)=&~P^{G}_{ES}(t)\,(1-b_{V}(t))\,b_{C}(t)+P^{G}_{L}(t),\notag \\
		P^{S}_{G}(t)=&~P^{ES}_{G}(t)\,(1-b_{C}(t))+P^{PV}_{G}(t).
	\end{align}
 With $C^{UG}_B(t)$ as the ToU tariff of purchasing power and $C^{S}_{UG}(t)$ the selling price for the external power exchanges, the cost $J_1(t)$ incurred by the microgrid at time instant $t$ is
\begin{align}
    \label{EQ_totalcost_01}
    J_1(t) = \,&~C^{UG}_B(t) P^{G}_{B}(t) b_{G}(t)-C^{S}_{UG}(t) P^{S}_{G}(t) (1-b_{G}(t)).	
\end{align}	
The internal power scheduling problem formulation for microgrids is discussed next.

\subsection{Problem Formulation}
\label{Sec_PrblmFormultn}

The microgrid internal power scheduling problem is modeled as a Mixed Integer Non-Linear Programming optimization with Predictive Scheduling (MINLP-PS). The non-linearity in the model mainly arises from the terms involving the multiplication of binary and power variables. Since solving a MINLP problem is computationally intensive, we use McCormick's relaxation \cite{mccormick_computability_1976} to convert the problem to a Mixed Integer Linear Programming with Predictive Scheduling (MILP-PS) \cite{dan_scenario_based_2020}. For the proposed Dynamic Internal Predictive Power Scheduling approach, the objective is transformed into a dynamic function by introducing a time-varying binary parameter. With this parameter, the objective function is no longer a static cost and provides for heterogeneity in power scheduling.

Converting the internal power scheduling problem at each instant into a Predictive Scheduling (PS) problem over a finite and receding time horizon provides an efficient solution to handle uncertainties \cite{turk_mpc_2021} and optimizes the performance of the ESS. With a prediction horizon of length $N_p$ at time epoch $T$, the objective is to minimize the costs incurred over this period. The local power generation and load demand for the prediction window are assumed to be known (based on real-world data). Efficient methods to predict these quantities, as discussed in the literature \cite{sun_review_2020}, can also be used to this end.

The optimization problem is subject to several constraints, namely the power balance constraints in \eqref{EQ_ELCons}, PV constraints in \eqref{EQ_SolarCons}, ESS SoC constraints in \eqref{EQ_Power_CH_DIS}--\eqref{EQ_DES_P_chmax}, and the external power exchange constraints in \eqref{EQ_UG_ex}. Thus, the power scheduling within the microgrid can be represented as a MINLP-PS problem as
	\begin{align}
		\label{eq:probform1}
		\mathcal{M}_1: & \underset{\begin{subarray}{c}
				P^{G}_{L}(t),\,P^{G}_{ES}(t),\,P^{ES}_{L}(t)\\
				P^{ES}_{G}(t),\,P^{PV}_{G}(t),\,P^{PV}_{ES}(t),\,P^{PV}_{L}(t)\\
				b_{C}(t),\,b_{G}(t),\,b_{V}(t)\\\end{subarray}}	{\operatorname{min}}\;\; \sum_{t=T+1}^{T+N_P} J_1(t) \\  \notag
		&\operatorname{constrained~by}\;\;\eqref{binVars_01},\eqref{EQ_ELCons}- \eqref{EQ_UG_ex}.
	\end{align}
 Here, $J_1(t)$ is the total cost \eqref{EQ_totalcost_01} incurred from power exchange with the external grid. This MINLP-PS is reformulated using McCormick's relaxation into a MILP-PS with the following substitutions
		\begin{align}
			\label{EQ_McC_var_cons}
			z^{G}_{ES}(t) & = P^{G}_{ES}(1-b_{V}(t)),\,\,
			y^{G}_{ES}(t) = z^{G}_{ES}(t) b_{C}(t),\notag \\
			w^{G}_{ES}(t) & = y^{G}_{ES}(t)  b_G (t);\notag \\
			w^{G}_{L}(t) & = P^{G}_{L} b_{G}(t);\,\,
			y^{ES}_{G}(t) = P^{ES}_{G}(t) (1-b_{C}(t)),\notag \\
			w^{ES}_{G}(t) & = y^{ES}_{G}(t) (1-b_{G}(t));\notag \\
			w^{PV}_{G}(t) &= P^{PV}_{G}(t) (1-b_{G} (t)),\notag \\
			y^{PV}_{ES}(t) & = P^{PV}_{ES}(t) b_{C}(t),\,\, w^{PV}_{ES}(t) = y^{PV}_{ES}(t) b_{V}(t),\notag \\
			w^{ES}_{L}(t) & = P^{ES}_{L}(t)(1-b_{C}(t)).
		\end{align}
The above substitutions linearize the constraints, resulting in \eqref{EQ_ELCons}--\eqref{EQ_SolarCons} being transformed into the following linear forms:
	\begin{align}
		\label{EQ_McC_PV_EL}
		P_{L}(t) = & ~w^{G}_{L}(t) + w^{ES}_{L}(t) + P^{PV}_{L}(t), \notag \\ 
		P^{PV}(t) = & ~w^{PV}_{G}(t) + w^{PV}_{ES}(t) + P^{PV}_{L}(t). 
	\end{align}

Similarly, the equations \eqref{EQ_Power_CH_DIS}--\eqref{EQ_DES_P_chmax} are modified into
		\begin{align}
			\label{EQ_McC_SoC}
			y_{ES}^{C}(t) = & ~w^{G}_{ES}(t)+w^{PV}_{ES}(t), \notag \\
			y^{ES}_{D}(t) = & ~w^{ES}_{G}(t)+w^{ES}_{L}(t), \notag \\
			SoC(h,t+1) = & ~SoC(t)\notag \\ +& (\eta_c y_{ES}^{G}(t) -\eta_d y^{ES}_{D}(t))/ES_{cap}. 
		\end{align}
 The modifications to the constraints for external power exchanges \eqref{EQ_UG_ex} are
	\begin{align}
		\label{EQ_McC_P_BS_US}
		y^{G}_{B}(t) &= w^{G}_{ES}(t)+P^{G}_{L}(t), \notag \\
		y^{S}_{G}(t) &= w^{ES}_{G}(t)+P^{PV}_{G}(t).
	\end{align}

Finally, the reformulated MILP-PS problem becomes
	\begin{align}
		\label{EQN_MILP_PS}
		\mathcal{M}^{S}_{MILP-PS}: & \underset{\begin{subarray}{c}
			P^{G}_{L}(t),\, w^{G}_{ES}(t),\, w^{ES}_{L}(t)\\
			 w^{ES}_{G}(t),\, w^{PV}_{G}(t),\,w^{PV}_{ES}(t),\,P^{PV}_{L}(t)\\\end{subarray}}	{\operatorname{min}}\;\; \sum_{t=T+1}^{T+N_P} J_{2}(t) \\  \notag
	&\operatorname{constrained~by}\;\;\eqref{binVars_01},\eqref{EQ_McC_PV_EL}- \eqref{EQ_McC_P_BS_US},
	\end{align}
where $J_{2}(t)= C^{UG}_B(t) y^{G}_{B}(t)-C^{S}_{UG}(t) y^{S}_{G}(t).$ Here, the MILP-PS optimization problem $\mathcal{M}^{S}_{MILP-PS}$ is referred to as the problem with a fixed objective function $J_{2}(t)$.

As previously mentioned, the MINLP-PS objective is further modified in the DIPPS approach. To this end, a time-varying binary parameter $b_{S}(t)$ is introduced as a multiplicative term with the power sold to the external grid. The resulting objective function is as follows:
\begin{align}
    J_D(t)= C^{UG}_B(t) y^{G}_{B}(t) - C^{S}_{UG}(t) y^{S}_{G}(t) - b_{S}(t)y^{S}_{G}(t).
\notag
\end{align}

This dynamic (time-varying) objective has the effect of maximizing profits from external power exchanges when the binary parameter is $0$. Meanwhile, with the binary parameter equal to $1$, the objective is to sell the power surplus stored in the ESS or the available surplus PV, irrespective of the ToU tariff rates. Thus, the binary parameter dynamically modifies the objective function, changing the internal power scheduling of the microgrid. The proposed DIPPS problem formulation can be expressed as
\begin{align}
    \label{EQN_MILP_PS_Dyn}
    \mathcal{M}^{D}_{MILP-PS}: & \underset{\begin{subarray}{c}
        P^{G}_{L}(t),\, w^{G}_{ES}(t),\, w^{ES}_{L}(t)\\
         w^{ES}_{G}(t),\, w^{PV}_{G}(t),\,w^{PV}_{ES}(t),\,P^{PV}_{L}(t)\\\end{subarray}}	{\operatorname{min}}\;\; \sum_{t=T+1}^{T+N_P} J_{D}(t) \\  \notag
&\operatorname{constrained~by}\;\;\eqref{binVars_01},\eqref{EQ_McC_PV_EL}- \eqref{EQ_McC_P_BS_US}.
\end{align}

The simulation setup for studying the performance of the DIPPS approach and the results for power scheduling are explained in the next section.

\section{Performance of Dynamic Internal Predictive Power Scheduling approach}
\label{Sec_Results}

In this paper, a real-world dataset (UCI Energy Dataset \cite{hebrail_uci_2012}) was used to generate the electrical load profiles of individual microgrids. This dataset contains the daily electrical load data measured between December 2006 and November 2010 (47 months) at a house in Sceaux, France. The PV production and ToU UG price data were obtained from \cite{hafiz_solar_2018}. The PV capacity was chosen to ensure that the total daily PV output matched the total load demand. The ESS storage capacity and the initial and final SoC were selected so that they could store the PV surplus and make it available in the absence of PV. The simulations were run in a MATLAB R$2020$b environment (Intel-i$7$ processor with a $16$ GB RAM). The optimization of the MINLP-PS and MILP-PS problems was solved using the Interior Point OPTimizer (IPOPT) solver from COIN-OR library in YALMIP \cite{lofberg_yalmip_2004}.

Three case studies were considered to evaluate the performance of the DIPPS approach. The electrical load data over a $24$-hour period, PV and ESS capacities were maintained the same for all three cases. In Case A, the static objective was considered, while Cases B and C utilized the dynamic objective with two different time-varying binary parameters to control the power sold to the external grid. In Case B, the time-varying parameter was set to $1$ between $0-6$ hours in the morning, while in Case C, the parameter was set to $1$ between $18-24$ hours.

\begin{figure*}[!htbp]
	\input{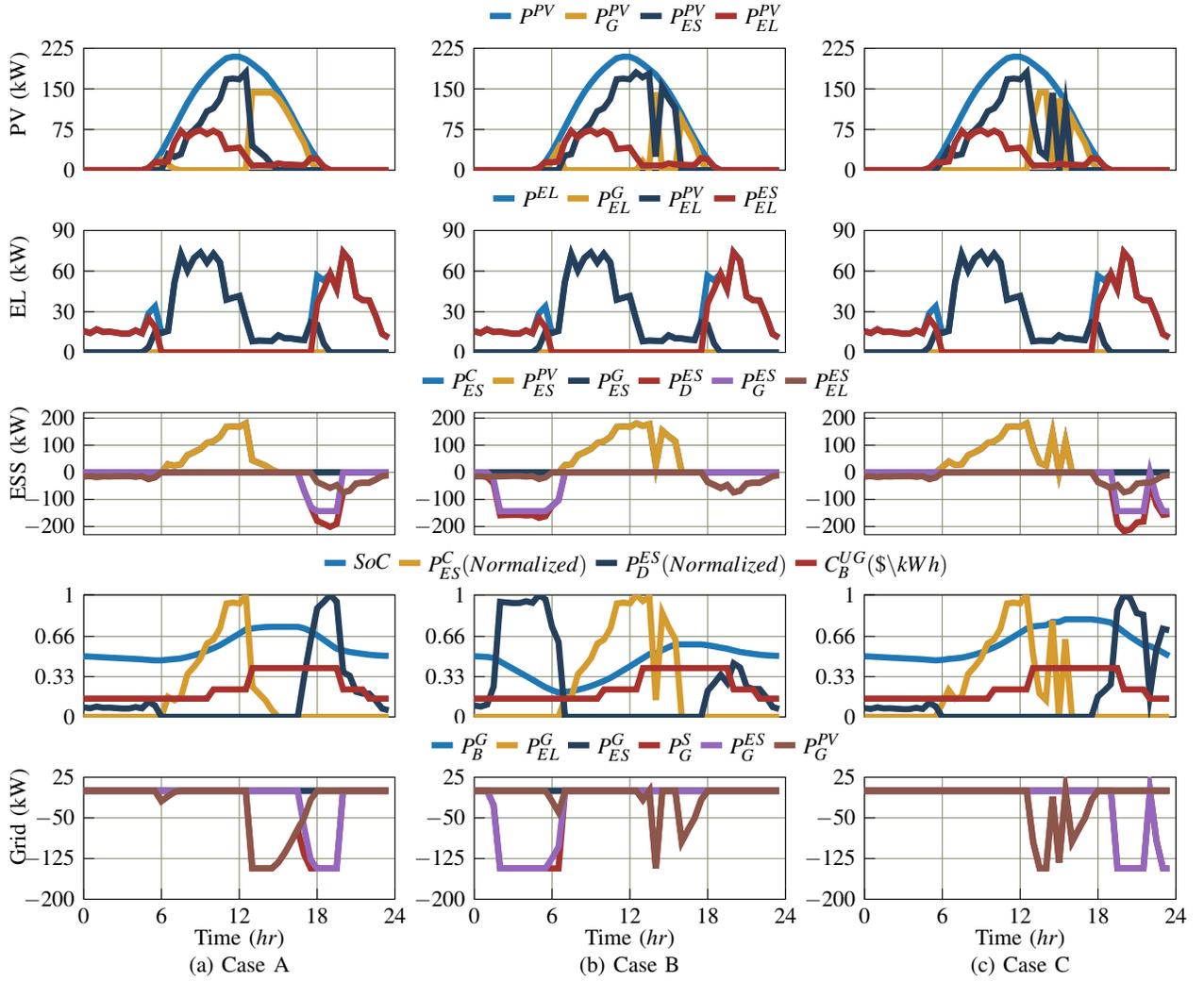}
	\caption{The results for internal power scheduling across three cases.}
	\label{IHS_PeX_comb}
\end{figure*}

Fig.~\ref{IHS_PeX_comb} provides detailed results of the individual variables related to internal power scheduling across the three cases. In Case A, illustrated in Fig.~\ref{IHS_PeX_comb}(a), the PV power primarily met the load, with any surplus being used to charge the ESS. The ESS then discharged to fulfill load requirements during periods when PV power was not available, particularly early morning and at night. The ESS charging occurred only under conditions of surplus PV power. After ensuring a final state of charge (SoC) of $0.5$ at the $24$-hour mark, any remaining stored power was sold during the high time-of-use (ToU) tariff period between $13-19$ hours. The fixed objective function ensured that the system relied primarily on the PV generation and the ESS, minimizing power purchases from the external grid. The time at which the power was sold to the external grid and the amount of power sold was influenced by the energy stored in the ESS and the ToU grid tariff rate.

For Case B, the results are presented in Fig.~\ref{IHS_PeX_comb}(b). The ESS discharged to meet the load until PV power generation started. Once PV power was available, it was used to satisfy the load, with any surplus stored in the ESS. The objective function, in this Case, was modified using a time-varying parameter set to $1$ between $0-6$ hours in the morning. This condition necessitated the ESS to discharge completely, reaching the minimum allowed SoC of $0.2$ by selling surplus energy to the grid after meeting the electrical load during this interval. For the remainder of the day, when the dynamic parameter was set to $0$, the system operated similarly to Case A, using PV power to meet the load and storing any excess. After $6$ hours, the power sold to the grid was solely from the surplus PV power, ensuring that the final SoC at the $24$-hour mark remained at $0.5$.

In Fig.~\ref{IHS_PeX_comb}(c), the results for Case C are presented. The time-varying parameter in the objective function was set to $1$ from $18-24$ hours in the evening. From $0-18$ hours, the system operated similarly to Case A, utilizing power stored in the ESS before PV generation began. When PV power was available, it was used to meet the electrical load, with any surplus stored in the ESS until the SoC limit of $0.8$ was reached. As in Case B, surplus PV power was sold to the grid between $12-18$ hours, but the SoC was ensured to reach $0.8$. During the $18-24$ hour period, the dynamic parameter required that power be sold to the grid while maintaining a SoC of $0.5$ at the end of the $24$ hours. In both Cases, B and C, the dynamic parameter in the objective function forced the sale of stored ESS power during the specified periods, regardless of the ToU tariff rates. As shown in Table~\ref{TABLE_TotGridPex}, the total power sold to the external grid over the simulated $24$ hours remained consistent across all three cases, with variations only in how the ESS was utilized to shift the timing of power exchanges in Cases B and C.

\begin{table}[htbp]
\centering
\caption{Total power sold to the external grid in the three cases and the power sold from the ESS}
\label{TABLE_TotGridPex}
\small
\begin{tabular}{ccr}
		\toprule
		& \multicolumn{2}{c}{Total power (kW)}
		\tabularnewline
		\cmidrule[\lightrulewidth](lr){2-3}
		{Cases}& $P^{S}_{G}$ & $P^{ES}_{G}$ 
		\tabularnewline
		\cmidrule[\lightrulewidth](lr){1-3}
        A & $1,850.39$ & $767.18$
        \tabularnewline
        B & $1,850.39$ & $1,400.18$
        \tabularnewline
        C & $1,850.39$ & $1,094.39$
        \tabularnewline
		\bottomrule
\end{tabular}
\end{table}

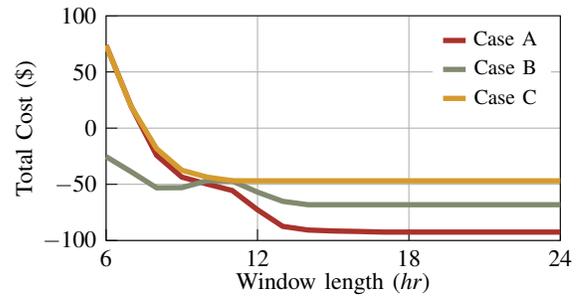
\begin{figure}[htbp]
	\centering
\begin{tikzpicture}

\definecolor{mycolor1}{RGB}{168,50,45}
\definecolor{mycolor2}{HTML}{818A6F}%
\definecolor{mycolor3}{HTML}{D59B2D}%
\definecolor{mycolor4}{HTML}{253F5B}%
\definecolor{mycolor5}{RGB}{136,176,197}
\definecolor{darkgray176}{RGB}{176,176,176}

\begin{axis}[
width=3in,
height=1.8in,
xmin=6, xmax=24,
ymin=-100, ymax=100,
tick align=inside,
tick pos=left,
x grid style={darkgray176},
xtick style={color=black},
y grid style={darkgray176},
ytick style={color=black},
xmajorgrids,
ymajorgrids,
xtick={6,12,18,24},
ytick={-100, -50, 0, 50, 100},
yticklabel style = {font=\small},
xticklabel style = {font=\small},
legend style={font=\footnotesize},
ylabel={\small Total Cost ($\$$)},
xlabel={\small Window length (\textit{hr})},
ylabel style={yshift=-1ex},
xlabel style={yshift=1ex},
legend style={
legend columns=1,
draw=none},
]
\addplot [line width=2.0pt, mycolor1]
table {%
6 73.7415491653824
7 19.0560015191341
8 -24.1342844757187
9 -43.5806910335917
10 -49.7607707562407
11 -55.5228988216829
12 -72.6664773837662
13 -87.4407640593058
14 -90.682028024907
15 -91.5307335051526
16 -91.958038120692
17 -92.5195325427392
18 -92.5195325427392
19 -92.5195325427392
20 -92.5195325427393
21 -92.5195325427393
22 -92.5195325427393
23 -92.5195325427393
24 -92.5195325427393
};
\addlegendentry{Case A}
\addplot [line width=2.0pt, mycolor2]
table {%
6 -25.3304224580579
7 -38.9370096130025
8 -53.1340325427393
9 -52.9740315369297
10 -46.8398464208143
11 -47.0628012029728
12 -56.9505733672377
13 -65.1299683548178
14 -68.1661004731667
15 -68.1661004731667
16 -68.1661004731667
17 -68.1661004731667
18 -68.1661004731667
19 -68.1661004731667
20 -68.1661004731667
21 -68.1661004731667
22 -68.1661004731667
23 -68.1661004731667
24 -68.1661004731667
};
\addlegendentry{Case B}
\addplot [line width=2.0pt, mycolor3]
table {%
6 73.7415491653824
7 19.0560015191341
8 -18.2536754566625
9 -37.4829752237612
10 -43.600818212496
11 -46.8191262875123
12 -46.9232927796844
13 -46.9232927796844
14 -46.9232927796844
15 -46.9232927796844
16 -46.9232927796844
17 -46.9232927796844
18 -46.9232927796844
19 -46.9232927796844
20 -46.9232927796844
21 -46.9232927796844
22 -46.9232927796844
23 -46.9232927796844
24 -46.9232927796844
};
\addlegendentry{Case C}
\end{axis}

\end{tikzpicture}
	\caption{Variation of the total cost of power exchange with the grid with an increase in predictive scheduling window.}
	\label{Fig_CostVnPart}
\end{figure}

Fig.~\ref{Fig_CostVnPart} shows how the total cost of power exchange with the grid varies when the predictive scheduling window length is increased from $6$ to $24$ hours for the three cases studied. For window length shorter than $6$ hrs, the ESS gets discharged to the limit and the solver is unable to meet the constraint of reaching the final SoC of $0.5$. By increasing the window length, the objective function can anticipate future electrical load, PV availability, and ToU tariff variations, thus enabling the ESS to be fully utilized. This allows surplus PV power, after storage, to be sold to the external grid without the need for external power purchases later in the day. A PS window of at least $6$ hours is recommended to maximize profit while maintaining the real-time computational feasibility. As demonstrated in the cases presented in Fig.~\ref{IHS_PeX_comb}, extending the scheduling window limits the charging-discharging cycles of the ESS to one per day, thereby prolonging the ESS lifespan. The maximum cost for a day is obtained for Case A, as the static objective ensures the power exchange with the grid corresponds to the highest ToU CG price (between $13-19$ hrs). For Cases B and C, the dynamic objective function forces the power exchanges to the predefined duration irrespective of the ToU price, leading to lower revenue margin. 

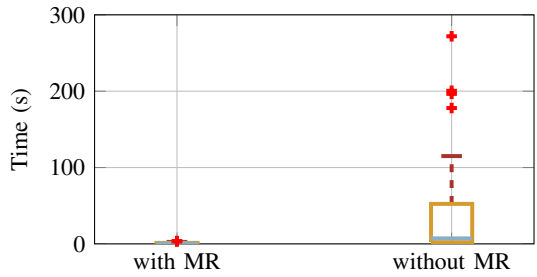
\begin{figure}[htbp]
    \begin{center}
        \definecolor{mycolor1}{RGB}{168,50,45}
\definecolor{mycolor2}{HTML}{818A6F}%
\definecolor{mycolor3}{HTML}{D59B2D}%
\definecolor{mycolor4}{HTML}{253F5B}%
\definecolor{mycolor5}{RGB}{136,176,197}

	\begin{tikzpicture}
		
\begin{axis}[%
width=2.3in,
height=1.2in,
scale only axis,
xmin=0.7,
xmax=2.3,
xtick={1,2},
yticklabel style = {font=\small,xshift=0ex},
xticklabel style = {font=\small,yshift=0ex},
xticklabels={\small {with MR},{without MR}},
ymin=0,
ymax=300,
ylabel style={font=\color{black}},
ylabel={\small Time (s)},
axis background/.style={fill=white},
xmajorgrids,
ymajorgrids
]
\addplot [color=mycolor1, dashed, line width=1.5pt, forget plot]
  table[row sep=crcr]{%
1	1.3365659\\
1	2.6308004\\
};
\addplot [color=mycolor1, dashed, line width=1.5pt, forget plot]
  table[row sep=crcr]{%
2	52.3967055\\
2	115.021364\\
};
\addplot [color=mycolor1, dashed, line width=1.5pt, forget plot]
  table[row sep=crcr]{%
1	0.2577223\\
1	0.3307947\\
};
\addplot [color=mycolor1, dashed, line width=1.5pt, forget plot]
  table[row sep=crcr]{%
2	0.7516883\\
2	2.25878815\\
};
\addplot [color=mycolor1, line width=1.5pt, forget plot]
  table[row sep=crcr]{%
0.9625	2.6308004\\
1.0375	2.6308004\\
};
\addplot [color=mycolor1, line width=1.5pt, forget plot]
  table[row sep=crcr]{%
1.9625	115.021364\\
2.0375	115.021364\\
};
\addplot [color=mycolor1, line width=1.5pt, forget plot]
  table[row sep=crcr]{%
0.9625	0.2577223\\
1.0375	0.2577223\\
};
\addplot [color=mycolor1, line width=1.5pt, forget plot]
  table[row sep=crcr]{%
1.9625	0.7516883\\
2.0375	0.7516883\\
};
\addplot [color=mycolor3, line width=1.5pt, forget plot]
  table[row sep=crcr]{%
0.925	0.3307947\\
0.925	1.3365659\\
1.075	1.3365659\\
1.075	0.3307947\\
0.925	0.3307947\\
};
\addplot [color=mycolor3, line width=1.5pt, forget plot]
  table[row sep=crcr]{%
1.925	2.25878815\\
1.925	52.3967055\\
2.075	52.3967055\\
2.075	2.25878815\\
1.925	2.25878815\\
};
\addplot [color=mycolor5, line width=1.5pt, forget plot]
  table[row sep=crcr]{%
0.925	0.4744511\\
1.075	0.4744511\\
};
\addplot [color=mycolor5, line width=1.5pt, forget plot]
  table[row sep=crcr]{%
1.925	7.2663259\\
2.075	7.2663259\\
};
\addplot [color=mycolor1, line width=1.5pt, only marks, mark=+, mark options={solid, draw=red}, forget plot]
  table[row sep=crcr]{%
1	3.2525721\\
1	3.648563\\
};
\addplot [color=mycolor1, line width=1.5pt, only marks, mark=+, mark options={solid, draw=red}, forget plot]
  table[row sep=crcr]{%
2	177.9270834\\
2	196.4787493\\
2	200.5048816\\
2	271.8731685\\
};
\end{axis}

\end{tikzpicture}%
        \vspace*{-2ex}
        \caption{Time taken to solve the problem in each window compared with and without using McCormick's relaxation}
        \label{Fig_MR_elapsed_comp}
    \end{center}
\end{figure}

Fig.~\ref{Fig_MR_elapsed_comp} compares the time required to solve each window in the MINLP-PS problem with and without the McCormick relaxation (MR). 
When compared over a single day, the mean time taken to solve the PS window with $6$ data points, the mean time to solve the MINLP-PS optimization is $38.27$s (with a min of $0.75$s, max of $271.87$s, and a standard deviation of $61.59$s), whereas, for the MILP-PS optimization, the mean time was reduced to $0.92$s (with a min of $0.26$s, max of $3.65$s, and a standard deviation of $0.81$s). This shows that solving the power scheduling as an MILP-PS problem reduces the mean computation time by $97.6\%$, and even the outliers with maximum time is reduced to $98.6\%$. Thus, McCormick relaxation improves the feasibility of solving power scheduling with MINLP-PS in real time.

\section{Conclusions}
\label{Sec_Conclusion}
A scalable Dynamic Internal Predictive Power Scheduling (DIPPS) approach for microgrid power management is proposed in this paper. DIPPS is designed to explore the external power exchange dynamics among heterogeneous prosumers. It utilizes a dynamic objective with a time-varying binary parameter to control the periods when power is exchanged with the external grid. This approach effectively leverages the available electrical storage to store surplus power and make it available for external exchanges at optimal periods. The computational complexity is significantly reduced by transforming the original MINLP-PS into a MILP-PS problem using McCormick's relaxation. We demonstrate the real-time implementation capability of DIPPS through simulations. For a predictive window window with $6$ data points, the average time to solve the MILP-PS formulation is $0.92$s compared to $38.27$s for the MINLP-PS formulation. This represents a $97.6\%$ reduction in computational speed. Longer predictive windows and reduced runtimes enable optimized charge-discharge cycles, enhancing the ESS lifespan. The effectiveness of the DIPPS approach is compared with a static objective using real-world data. Three cases are analyzed, each with the same PV, ESS capacities, and load but with time-varying parameters. This validates the scheduling approach's ability to optimize power exchanges with the grid while optimally utilizing available distributed resources.

\bibliographystyle{IEEEtran}

\end{document}